\documentclass[runningheads]{llncs}
\usepackage[T1]{fontenc}
\usepackage{graphicx}
\usepackage{xspace} 
\usepackage{xcolor} 
\usepackage[dvipsnames]{xcolor} 
\usepackage{hyperref} 
\usepackage{cleveref} 
\usepackage{listings} 
\usepackage{apxproof} 
\usepackage{makecell} 
\usepackage{orcidlink} 

\def\grind{\textsc{grind}\xspace}
\def\querysmt{\textsc{querysmt}\xspace}
\def\querysmtminus{\textsc{querysmt}$-$\xspace}
\def\cvc5{\texttt{cvc5}\xspace}
\def\skolemizeAll{\texttt{skolemizeAll}\xspace}
\def\leansmt{\textsc{lean-smt}\xspace}
\def\duper{\textsc{duper}\xspace}
\def\aesop{\textsc{aesop}\xspace}
\def\metis{\textsc{metis}\xspace}
\def\leanauto{\textsc{lean-auto}\xspace}

\definecolor{keywordcolor}{rgb}{0.7, 0.1, 0.1}   
\definecolor{tacticcolor}{rgb}{0.0, 0.1, 0.6}    
\definecolor{commentcolor}{rgb}{0.4, 0.4, 0.4}   
\definecolor{symbolcolor}{rgb}{0.0, 0.1, 0.6}    
\definecolor{sortcolor}{rgb}{0.1, 0.5, 0.1}      
\definecolor{attributecolor}{rgb}{0.7, 0.1, 0.1} 

\begin{document}
\title{Hint-Based SMT Proof Reconstruction}
%
%
\author{Joshua Clune\inst{1} \orcidlink{0000-0003-4047-6196} \and
Haniel Barbosa\inst{2} \orcidlink{0000-0003-0188-2300} \and
Jeremy Avigad\inst{1} \orcidlink{0000-0003-1275-315X}}
\authorrunning{J. Clune et al.}
\institute{Carnegie Mellon University, Pittsburgh, PA, USA\\
\and
Universidade Federal de Minas Gerais, Belo Horizonte, Brazil}
\maketitle              
\begin{abstract}
There are several paradigms for integrating interactive and automated theorem provers, combining the convenience of powerful automation with strong soundness guarantees.
We introduce a new approach for reconstructing proofs found by SMT solvers which we intend to be complementary with existing techniques.
Rather than verifying or replaying a full proof produced by the SMT solver, or
at the other extreme, rediscovering the solver's proof from just the set of
premises it uses, we explore an approach which helps guide an interactive
theorem prover's internal automation by leveraging derived facts during solving,
which we call hints.
This makes it possible to extract more information from the SMT solver's proof without the cost of retaining a dependency on the SMT solver itself. We implement a tactic in the Lean proof assistant, called \querysmt, which leverages hints from the \cvc5 SMT solver to improve existing Lean automation. We evaluate \querysmt's performance on relevant Lean benchmarks, compare it to other tools available in Lean relating to SMT solving, and show that the hints generated by \cvc5 produce a clear improvement in existing automation's performance.

\end{abstract}

\section{Introduction}

When it comes to formal verification, interactive and automatic theorem provers
(ITPs or proof assistants, and ATPs, respectively) have complementary strengths.
Proof assistants offer powerful languages for expressing arbitrary mathematical
statements and the ability to verify claims down to domain-specific axioms and the rules of the underlying logical foundation, but doing so often requires considerable effort. Automatic provers offer push-button verification but often fail to scale to complex
verification tasks and do not provide the same strong guarantees as interactive theorem provers.
Proof assistants like Isabelle/HOL~\cite{DBLP:conf/cade/BlanchetteBP11,Blanchette2016-qed,Lachnitt2025,Schurr2021}, Rocq~\cite{Armand2011,Czajka2018}, and
Lean~\cite{Bving2025,Mohamed2025,DBLP:conf/itp/NormanA25} aim for the
best of both worlds by translating goals in a proof assistant to the language of
a powerful external prover and then using information from the external prover
to reconstruct a proof of the original result that is checked within the ITP.
The challenge, then, is to bridge the gap and establish appropriate communication between the two.

There are several existing paradigms for effectively communicating between ATPs and proof assistants. Their approaches to reconstructing ATP proofs vary. One approach, often used for superposition theorem provers such as \texttt{E}~\cite{Vukmirovi2023}, \texttt{Vampire}~\cite{Kovcs2013} and \texttt{Zipperposition}~\cite{Vukmirovi2021}, is to supply a large number of premises to the ATP, use the ATP's proof to identify a minimal subset of necessary premises, and then supply just those premises to internal ITP automation such as \metis~\cite{hurd2003d}. This has the benefit of entirely removing the call to the external prover, but has the possibility of failure because it ultimately depends on internal automation independently discovering a proof.

SMT solvers~\cite{Barrett2021}, which combine generic first-order reasoning with theory-specific decision procedures, generally warrant reconstruction methods which more closely follow the external solver's original proof. Isabelle/HOL relies on internal tactics to replay each step in the certificates from the SMT solvers it supports (\texttt{z3}~\cite{deMoura2008-z3}, \texttt{veriT}~\cite{Bouton2009}, and \cvc5~\cite{Barbosa2022-cvc5}). SMTCoq uses the SMT solver \texttt{veriT}, which can produce detailed certificates, and checks its certificates via a formally verified procedure within Rocq. \leansmt uses a mixtures of these two approaches, but mostly proof replay, to reconstruct certificates from \cvc5.

Our goal in this paper is to explore an alternative approach to proof reconstruction. Instead of reconstructing the SMT solver's proof exactly, retaining a dependency on it, or discarding all information about the solver's proof except the set of premises used, our approach uses the SMT solver's proof to help guide ITP automation by providing \emph{hints}.

We develop a Lean tactic, called \querysmt, which uses \leanauto~\cite{qian2025lean-auto} to export problems from Lean's dependent type theory to the language of an SMT solver. We then instrument \cvc5 to report the preprocessing and theory reasoning performed while solving the translated problem. This is used by \querysmt to insert, in the Lean source file, a self-contained proof script for the goal, with Lean formulations of the theory-specific facts formalized as subgoals. The proof script uses \grind, a built-in Lean tactic inspired by SMT solvers, to supply proofs of those subgoals, and uses a proof-producing superposition prover, \duper~\cite{CluneEtAl2024Duper}, to prove the original goal using those facts. The result is a structured proof that users can inspect, modify, and simplify, if they wish. Notably, the proof does not depend on calling an SMT solver anymore.
As in Isabelle's Sledgehammer when not using the \textsf{smt} tactic, the call to the external prover disappears.

We believe this approach has complementary strengths to others. SMT solvers are
notoriously unstable; small changes in context, even as minor as renaming
variables, can cause proofs to break, as well as different solver versions running
on the same problem~\cite{Zhou2023}.
Therefore, eliminating the SMT call in favor of a modular source-code proof
results in a more stable artifact. Additionally, we consider it a benefit that users can inspect and modify the resulting proof script. Powerful automation may be good at finding a
proof, but it rarely yields the nicest one.

We demonstrate the method with arithmetic and inductive types, two of the most important and common theories in Lean and other proof assistants, though our approach is not specific to these. We evaluate the approach on relevant benchmark problems from Lean's Init, Batteries, and Mathlib \cite{ThemathlibCommunity2020} libraries, and compare it to other SMT-related tools available in Lean. We show that although proof reconstruction does not always succeed, incorporating SMT hints significantly improves internal automation's chances of successful reconstruction.

Our contributions are as follows:
\begin{itemize}
\item We augment \cvc5 with the ability to record data that will be useful for proof reconstruction and report it back to Lean.
\item We develop a method of translating statements about natural numbers to SMT queries on the integers.
\item We augment our back-end reconstruction, \duper \cite{CluneEtAl2024Duper}, to implement a set of support strategy \cite{LPAR-21S:Set_Support_Theory_Reasoning}, improving its ability to incorporate SMT hints.
\item We evaluate \querysmt's performance on Lean benchmarks, showing that \querysmt compares favorably to existing SMT-related Lean automation and that SMT hints produce a clear improvement in \duper's performance.
\end{itemize}


\section{\querysmt Overview}\label{sec:querysmt-overview}

\begin{figure}[ht]
    \centering
    \includegraphics[width=0.75\linewidth]{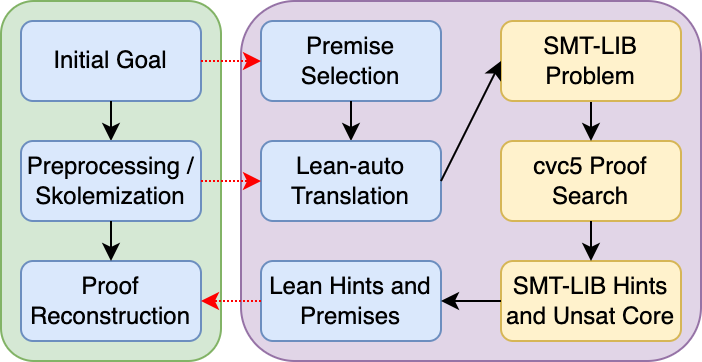}
    \caption{Overview of the \querysmt tactic. Blue boxes indicate Lean stages. Yellow boxes indicate SMT stages. Stages in the green area directly transform the Lean goal and must be replayed in the final suggested proof script. Stages in the purple area do not transform the goal and therefore do not need to be replayed. Red dotted lines indicate information transfer between stages.}
    \label{fig:querysmt-outline}
\end{figure}

To demonstrate and evaluate our approach, we develop \querysmt, a Lean tactic which utilizes hints from the \cvc5 SMT solver to suggest a self-contained proof script.
The overall structure of \querysmt is given in \Cref{fig:querysmt-outline}.

\vspace{-2ex}

\querysmt consists of five primary components: preprocessing, translation, hint generation, hint interpretation, and proof reconstruction. Preprocessing transforms the goal into a form that reduces the likelihood of \cvc5 producing hints that can't be interpreted or proven. Translation from Lean's dependent type theory \cite{Theorem_Proving_in_Lean4} to the many-sorted first-order logic used by SMT solvers \cite{BarFT-RR-17} is handled by \leanauto \cite{qian2025lean-auto}, with extensions made to \leanauto's SMT-LIB translation described in \Cref{sec:translation}. Hint generation consists of recording facts generated over the course of \cvc5's proof search and identifying an unsat core from the given goal and set of premises provided by premise selection. Hint interpretation involves translating \cvc5's generated hints into usable Lean expressions. Finally, proof reconstruction accepts \cvc5's unsat core and the output of hint interpretation and uses them to suggest a self-contained proof script that the user can examine and modify.

\vspace{-4ex}
\begin{figure}[ht]
\begin{minipage}{0.95\textwidth}\vspace{0pt}
\begin{lstlisting}[language=lean]
example (f : Int → Int) (h1 : ∀ x y, f x = f y → x = y)
  (h2 : ∃ x, ∀ y, f x ≤ f y) : ∃ x, ∀ y, x ≠ y → f x < f y := by
  apply @Classical.byContradiction
  intro negGoal
  skolemizeAll
  have smtLemma0 : (∀ (_i_0 : Int), f sk0 ≤ f _i_0) →
    ∀ (_i_0 : Int), ¬f sk0 + -Int.ofNat 1 * f _i_0 ≥ Int.ofNat 1 :=
    by grind
  have smtLemma1 :
    (∀ (_i : Int),
        have _let_1 := sk1 _i;
        ¬(¬_i = _let_1 → f _i < f _let_1)) →
      ∀ (BOUND_VARIABLE_4962 : Int), f BOUND_VARIABLE_4962 + -Int.ofNat 1 * f (sk1 BOUND_VARIABLE_4962) ≥ Int.ofNat 0 :=
    by grind
  have smtLemma2 :
    have _let_1 := f (sk1 sk0);
    have _let_2 := f sk0;
    have _let_3 := _let_2 + -Int.ofNat 1 * _let_1;
    (¬_let_3 ≥ Int.ofNat 0 ∨ _let_2 = _let_1) ∨
    _let_3 ≥ Int.ofNat 1 := by grind
  duper [h1, h2, negGoal, smtLemma0, smtLemma1, smtLemma2] []
\end{lstlisting}
\end{minipage}
\caption{A proof script suggested by \querysmt. In this example, the \skolemizeAll call produces \texttt{(sk0 : Int)} and \texttt{(sk1 : Int $\rightarrow$ Int)} from \texttt{h2} and \texttt{negGoal} respectively.}
\label{fig:proof-script-example}
\end{figure}
\vspace{-2ex}

\Cref{fig:proof-script-example} contains an example of a proof script suggested by \querysmt. The first three lines reproduce \querysmt's preprocessing and Skolemization, the \texttt{have} statements that follow are Lean translations of hints recommended by \cvc5, and the final line is a call to \duper which uses the recommended hints to complete the overall goal. As of Lean version \texttt{v4.22.0}, neither \grind nor \duper can prove this example alone, as \duper cannot verify the hints output by \cvc5 and \grind cannot complete the overall proof even with \cvc5's hints. But by using \grind to reconstruct \cvc5's theory reasoning and \duper to reconstruct \cvc5's logical reasoning, \querysmt is able to complete the proof.

On a first pass, the proof script in \Cref{fig:proof-script-example} is not very easy to read. \querysmt does some work for the user by filtering the set of suggested hints to only include those necessary for \duper's final proof,\footnote{In the example from \Cref{fig:proof-script-example}, \cvc5 generates 8 hints initially, but \querysmt is able to filter them down to 3 for the final proof script.} but even so, the individual hints can still be unnecessarily verbose. In this example, the hypotheses in \texttt{smtLemma0}'s and \texttt{smtLemma1}'s implications directly match the statements of \texttt{h2} and \texttt{negGoal} after the modifications made by \texttt{skolemizeAll}, making the hypotheses redundant. The proof script can be made significantly more readable by removing these hypotheses and making some other minor changes, such as renaming \texttt{BOUND\_VARIABLE\_4962} as \texttt{z}, performing zeta-reduction to remove unnecessary \texttt{let} expressions, and simplifying expressions of the form \texttt{-Int.ofNat 1 * $e$} to \texttt{-$e$}. A simplified version of \Cref{fig:proof-script-example}'s proof script is shown in \Cref{fig:proof-script-example-modified}. \querysmt does not currently have a postprocessing module to perform such simplifications automatically, but this would be worth exploring as future work.

\vspace{-4ex}
\begin{figure}[ht]
\begin{minipage}{0.95\textwidth}
\begin{lstlisting}[language=lean]
example (f : Int → Int) (h1 : ∀ x y, f x = f y → x = y)
  (h2 : ∃ x, ∀ y, f x ≤ f y) : ∃ x, ∀ y, x ≠ y → f x < f y := by
  apply @Classical.byContradiction
  intro negGoal
  skolemizeAll
  have smtLemma0 : ∀ (_i_0 : Int), ¬f sk0 - f _i_0 ≥ 1 := by grind
  have smtLemma1 : ∀ (z : Int), f z - f (sk1 z) ≥ 0 := by grind
  have smtLemma2 :
    (¬f sk0 - f (sk1 sk0) ≥ 0 ∨ f sk0 = f (sk1 sk0)) ∨
    f sk0 - f (sk1 sk0) ≥ 1 := by grind
  duper [h1, h2, negGoal, smtLemma0, smtLemma1, smtLemma2] []
\end{lstlisting}
\end{minipage}
\caption{A simplified version of the proof script shown in \Cref{fig:proof-script-example}.}
\label{fig:proof-script-example-modified}
\end{figure}
\vspace{-4ex}
\section{Preprocessing}
\label{sec:preprocessing}

Preprocessing takes a Lean goal of the form $\Gamma \vdash p : \textsf{Prop}$ and transforms it into a goal of the form $\Gamma' \vdash \textsf{False} : \textsf{Prop}$ where all hypotheses in $\Gamma'$ are Skolemized and $\Gamma'$ entails $\Gamma$ and $\neg p$. This transformation is necessary for two reasons.

First, the hints that \cvc5 generates may depend on the falsity of the initial target $p$. When this occurs, the generated hints are not entailed by $\Gamma$, meaning that any proof automation that attempts to derive the hint from $\Gamma$ is doomed to fail. By preprocessing the goal into a state where the local context $\Gamma'$ entails both $\Gamma$ and $\neg p$, \querysmt ensures that when it comes time to prove the hints provided by \cvc5, all pertinent information is accessible in the local context.

Second, the hints that \cvc5 generates may include constants that do not appear
in the original SMT-LIB problem produced by \leanauto. This occurs when \cvc5
internally Skolemizes existential quantifiers, generating constants of the form
\texttt{@QUANTIFIERS\_SKOLEMIZE\_X}.\footnote{As defined in
  \url{https://cvc5.github.io/docs/cvc5-1.3.1/skolem-ids.html}, this constant
  corresponds to a term resulting from Skolemization, which could be defined
  e.g.\ via Hilbert's choice operator if the Skolemized quantifier were given as
  well. However, this is only done when proofs are generated.}
While it is possible to recover the meaning of these constants by following the
chain of inferences that were taken to reach \cvc5's Skolemization inferences, this approach essentially amounts to partial proof replay, which goes against the intention of \querysmt's design.
To avoid this issue, \querysmt handles Skolemization in Lean prior to calling \leanauto's translation procedure. We define a tactic called \skolemizeAll which iterates through every hypothesis in the local context and attempts to remove existential quantifiers and negated universal quantifiers, replacing them with fresh free variables. When no Skolemization is necessary because neither the original context $\Gamma$ nor the negated target $\neg p$ contain existential quantifiers or negated universal quantifiers, \querysmt notes that \skolemizeAll has no impact on the goal state and omits it from the final proof script suggestion.

We note that because Lean's dependent type theory includes empty types, \skolemizeAll may fail in cases where it is unable to verify type inhabitation. For example, Skolemizing the hypothesis $h : (\exists x : \alpha, P\ x) \lor \texttt{True}$ requires generating a free variable $y : \alpha$ and replacing $h$ with $h' : P\ y \lor \texttt{True}$. However, $h$ alone does not entail that $\alpha$ is inhabited. Unless $\alpha$ is already known to be inhabited, \skolemizeAll cannot soundly add $y : \alpha$ to the local context. Consequently, there are some theorems that \querysmt is fundamentally unable to tackle due to the presence of empty types or possibly empty polymorphic types.
This limitation is inherent to approaches translating from Lean's logic to
SMT-LIB, and also affects e.g.\ \leansmt~\cite[Sec. 3.2]{Mohamed2025}, where
proof replay may fail when the solver relied on the non-emptiness of a type
and that cannot be established by Lean.


\section{Translation to SMT-LIB} \label{sec:translation}

\subsection{\leanauto}

For translation to the many-sorted first-order logic of SMT-LIB, we rely on \leanauto~\cite{qian2025lean-auto}. \leanauto normalizes universe levels, monomorphizes definitions, handles definitional equalities, and broadly takes care of the many features that make Lean's type theory complex.
When targeting SMT-LIB, \leanauto attempts to translate Lean types to their closest SMT-LIB analogues. For example, Lean's \texttt{Prop} and \texttt{Bool} types are both translated to SMT-LIB's \texttt{Bool} sort, and Lean's \texttt{Int} and \texttt{Nat} types are both translated to SMT-LIB's \texttt{Int} sort.
This ensures that the translation takes full advantage of SMT solvers' theories, but sometimes creates complications for interpreting
the SMT hints (see \Cref{sec:hint-interpretation}).

\leanauto primarily translates essentially higher-order problems in Lean to monomorphic higher-order logic, but also can translate essentially first-order problems in Lean to first-order logic for the purpose of targeting SMT-LIB \cite{CluneEtAl2024Duper,qian2025lean-auto}. Although the most recent version of the SMT-LIB standard (Version 2.7) defines an SMT-LIB logic which extends beyond the many-sorted first-order logic adopted by the previous version \cite{BarFT-RR-25}, this SMT-LIB logic is not specifically targeted by \leanauto. When the upcoming Version 3 of the SMT-LIB standard is released with a higher-order base logic, we expect it will be fruitful to modify \leanauto so that it uses its primary translation procedure even when targeting SMT-LIB. This would expand the fragment of Lean that can be effectively translated to SMT-LIB, but is beyond the scope of this paper.

\subsection{Translating Natural Numbers}\label{sec:naturals-translation}

Unlike the translation procedures used by SMTCoq \cite{Armand2011} or Isabelle's Sledgehammer \cite{MengP08,MengP09,PaulsonB10,PaulsonS07}, \leanauto does not adopt an encoding-based approach to translating natural numbers. Instead, \leanauto directly translates Lean terms of type \texttt{Nat} to SMT-LIB terms of sort \texttt{Int} along with assertions of the form \texttt{(assert (>= n 0))}. Although this approach is sufficient for many common use cases, it is incomplete and can lead to unsound translations when natural numbers are embedded in larger structures or inductive types. For example, \leanauto is able to soundly translate $\texttt{example}\ (n : \texttt{Nat}) : 0 \le n := \ldots$ into an unsatisfiable SMT-LIB problem, but the SMT-LIB problem generated from $\texttt{example}\ (x : \texttt{Nat} \times \texttt{Nat}) : 0 \le x.\texttt{fst} := \ldots$ is satisfiable because \leanauto fails to assert that both projections of $x$ must be nonnegative.

We extend \leanauto's procedure by preserving the direct translation from Lean \texttt{Nat} terms to SMT-LIB \texttt{Int} terms while expanding the set of circumstances in which nonnegativity assertions are made. For every Lean type $\alpha$ which appears in the problem, we define an SMT-LIB predicate $\texttt{wf}_\alpha : \hat{\alpha} \rightarrow \texttt{Bool}$ where $\hat{\alpha}$ is the SMT-LIB sort corresponding to $\alpha$. This predicate is meant to encode the fact that the SMT-LIB term it applies to is \emph{well-formed}, meaning it satisfies all of the nonnegativity constraints imposed by the \texttt{Nat} type on the Lean term from which it was derived. As a concrete example, if $\alpha = \texttt{Nat} \times \texttt{Int}$, then $\texttt{wf}_\alpha$ asserts that the first projection of the term it applies to is nonnegative.

\begin{definition} \label{def:wf}
The predicate $\textup{\texttt{wf}}_\alpha\ \textup{\texttt{x}}$ is defined inductively on $\alpha$ as follows:
\begin{itemize}
    \item If $\alpha = \textup{\texttt{Nat}}$ then:
    \begin{itemize}
        \item \textup{$\texttt{wf}_\alpha\ \texttt{x} = \texttt{(>= x 0)}$}
    \end{itemize}
    \item If $\alpha = \alpha_1 \rightarrow \alpha_2$ then:
    \begin{itemize}
        \item \textup{$\texttt{wf}_\alpha\ \texttt{x} = \texttt{(forall ((y $\hat{\alpha_1}$)) (=> (wf$_{\alpha_1}$ y) (wf$_{\alpha_2}$ (x y))))}$}
    \end{itemize}
    \item If $\alpha$ is a structure\footnote{In Lean's type theory, all structures are inductive datatypes, meaning there is no need to distinguish between them \cite{Theorem_Proving_in_Lean4}. \Cref{def:wf}'s treatment of structures is logically equivalent to its treatment of inductive datatypes with one constructor, so it would be straightforward to eliminate the distinction. We nonetheless distinguish between them because only structures are guaranteed to have projections already defined in Lean. This has consequences for hint interpretation, which are discussed in \Cref{sec:hint-interpretation}.} with projections $p_1 : \alpha \rightarrow \alpha_1, \ldots p_n : \alpha \rightarrow \alpha_n$ then:
    \begin{itemize}
        \item \textup{$\texttt{wf}_\alpha\ \texttt{x} = \bigwedge\limits_{i=1}^n \texttt{wf}_{\alpha_i}\ \texttt{($\hat{p_i}$\ \texttt{x})}$}
    \end{itemize}
    \item If $\alpha$ is an inductive datatype with constructors $(c_1 : \beta_{1, 1} \rightarrow \ldots \rightarrow \beta_{1, m_1} \rightarrow \alpha), \ldots (c_n : \beta_{n, 1} \rightarrow \ldots \beta_{n, m_n} \rightarrow \alpha)$ and is translated to a datatype with constructors $\hat{c_1}, \ldots \hat{c_n}$ and selectors $s_{i, j} : \hat{\alpha} \rightarrow {\hat{\beta_{i, j}}}$ then:
    \begin{itemize}
        \item \textup{$\texttt{wf}_\alpha\ \texttt{x} = \bigwedge\limits_{i=1}^n \texttt{(=> (}\texttt{x}\ \texttt{is}\ \hat{c_i}\texttt{)}\ \texttt{(}\bigwedge\limits_{j=1}^{m_i} \texttt{wf}_{\beta_{i, j}}\ \texttt{(}s_{i, j}\ \texttt{x} \texttt{)))}$}
    \end{itemize}
    \item Otherwise:
    \begin{itemize}
        \item \textup{$\texttt{wf}_\alpha\ \texttt{x} = \texttt{True}$}
    \end{itemize}
\end{itemize}
\end{definition}

To ensure that the semantics of the SMT-LIB problem coincide with the semantics of the original Lean goal, \texttt{wf} constraints are inserted such that almost all terms which appear in the SMT-LIB problem are provably well-formed. Whenever an SMT-LIB function or constant is declared, an assertion is added to guarantee that it is well-formed. Additionally, whenever an SMT-LIB formula is translated from a Lean proposition, the translation of quantifiers is modified to assert the well-formedness of the introduced variable. Universal quantification over $\alpha$ is translated to \texttt{(forall ((x $\hat\alpha$)) (=> (wf$_\alpha$ x) ($\ldots$)))} and existential quantification over $\alpha$ is translated to \texttt{(exists ((x $\hat\alpha$)) (and (wf$_\alpha$ x) ($\ldots$)))}.

We note that this approach to translating natural numbers appears to coincide with \texttt{Trakt}'s methodology for handling partial embeddings on problems that do not involve inductive datatypes \cite{Blot2023}. Our approach diverges from \texttt{Trakt}'s when inductive datatypes are involved because \texttt{Trakt} is intended to provide preprocessing transformations that are independent of the targeted backend while we explicitly aim to take advantage of SMT solvers' built-in datatype support.

\begin{theorem} \label{thm:well-formedness}
    All terms except datatype selectors\footnote{\Cref{thm:well-formedness} does not assert that datatype selectors are well-formed because in general they aren't. When a selector is passed a well-formed datatype built from the wrong constructor, the resulting application's output is only constrained by its sort (see remark 20 of the SMT-LIB standard \cite{BarFT-RR-25}). Therefore, the output may fail to satisfy the nonnegativity constraints required by well-formedness.} which appear in an SMT-LIB problem generated by our procedure are provably well-formed.
\end{theorem}

\begin{proofsketch}
    Let $t : \hat{\alpha}$ be some term which appears in the generated SMT-LIB problem. The proof proceeds by induction on $t$. Here, we show one nontrivial case. For a proof sketch which covers more cases, see \Cref{pf:well-formedness}.
    \begin{itemize}
        \item If $t$ is an application of of the form \texttt{($t_1$ $t_2$)}, then the Lean term corresponding to $t_1$ has type $\beta \rightarrow \alpha$ and the Lean term corresponding to $t_2$ has type $\beta$.\footnote{We can infer the form of the Lean term corresponding to $t$ because Lean-auto's monomorphization procedure preserves term structures during translation.} By the inductive hypothesis, $t_2$ is well-formed and $t_1$ is either well-formed or a selector function.
        \begin{itemize}
            \item If $t_1$ is well-formed, then from the definition of well-formedness on functions, \texttt{wf$_{\beta \rightarrow \alpha}$ $t_1$ $=$ (forall ((y $\hat{\beta}$)) (=> (wf$_\beta$ y) (wf$_\alpha$ ($t_1$ y))))}.\\
            From this and \texttt{wf$_\beta$ $t_2$}, it follows that \texttt{wf$_\alpha$ ($t_1$ $t_2$)} as desired.
            \item If $t_1$ is a selector function, then $\beta$ is a structure or inductive datatype and $t_1$ has some associated constructor $\hat{c}$. From the definition of well-formedness on structures and inductive datatypes, \texttt{wf$_\beta$ $t_2$} entails \texttt{(=> ($t_2$ is $\hat{c}$) (wf$_\alpha$ ($t_1$ $t_2$)))}. \leanauto's translation procedure guarantees that if \texttt{($t_1$ $t_2$)} appears in the generated SMT-LIB problem, then $t_2$ satisfies the tester for $t_1$'s constructor, meaning \texttt{($t_2$ is $\hat{c}$)} holds. From this and the implication entailed by \texttt{wf$_\beta$ $t_2$}, it follows that \texttt{wf$_\alpha$ ($t_1$ $t_2$)}.
        \end{itemize}
    \end{itemize}
\end{proofsketch}


\section{Hint Generation}
\label{sec:hint-generation}

We have instrumented the \cvc5 SMT solver to report hints for external tools
(such as \querysmt) based on logical consequences derived during proof search
from the input formula and the theories supported by the solver.

The hints are collected from the internal proof produced by
\cvc5~\cite{Barbosa2022}, which is then discarded.
We consider three kinds of hints: \emph{preprocessing lemmas}, \emph{theory
  lemmas}, and \emph{rewrite steps}. We consider them because they each contain theory reasoning performed by the solver while proving the given
goal.
We restrict ourselves to hints that were \emph{useful} to the solver,
i.e., they were used in the final proof.
A useful by-product of the search for hints is to also collect an \emph{unsat
  core} of the input, i.e., the elements of the input that were present
in the proof.

Preprocessing is a key element of SMT solving where an input formula is
simplified according to a series of preprocessing passes, each potentially
modifying the input, be it by replacing the formula with a simplified version or
generating new formulas entailed by it.
The hints collected are entailments between the input formula and the
preprocessed ones. An example of a key preprocessing step performed by \cvc5 is the inference and application of a substitution
over part of the input to eliminate terms that are definable by others, as per
the rest of the input (e.g.\ if the input contains the equality $x = 5$, a
substitution corresponding to $x \mapsto 5$ is applied to the rest of the input
to remove $x$~\cite[Sect. 5.1]{Barbosa2022}).

Theory lemmas are valid disjunctions of literals from one or more theories.
They generally correspond to explanations of why a given assignment of truth
values to literals is inconsistent (e.g.\ assigning \texttt{True} to both
$a = b$ and $f(a)\neq f(b)$), and allow the pruning of search space relative to
that wrong assignment.

Finally, we also collect intermediate theory reasoning steps applied during
preprocessing and during theory lemma generation that correspond to rewrite
steps.
These are important because they encapsulate key theory reasoning that would be
difficult to gauge from just the preprocessing or theory lemmas themselves.
An example is how \cvc5 reduces all arithmetic terms to sums of monomials.
Since the hints will contain only the fully reduced terms, including the
intermediate rewrite steps increases the information made available to \querysmt.

While the proof contains other elements (besides the justification themselves for those hints), such as the resolution reasoning performed by a SAT solver on the Boolean structure of the formula and its connection with the hints, we do not consider them because this logical reasoning can be performed by \duper.

\paragraph{Normalization of AC operators.} Internally \cvc5, as most SMT
solvers, represent associative operators not as binary but as $n$-ary operators,
so it is common to apply a ``flattening'' simplification so that
terms such as \texttt{(* (* $x_1$ $x_2$) $x_3$)} and \texttt{(* $x_2$ (* $x_3$ $x_1$)))} are represented as the
same term \texttt{(* $x_i$ $x_j$ $x_k$)}. Not only are applications of the operator
flattened, but arguments are rearranged according to a canonical order.
This normalization is useful for solving, but it complicates proof
reconstruction in systems not representing AC operators in a normalized form.
Since hints generated by \cvc5 refer to the normalized version of a term,
and this normalization is not present in the hints (it would only be present in
a proper proof), the applicability of the hints by \querysmt can be limited
unless \querysmt can infer the relation between original terms and their normalized versions.
A potential solution is to integrate facts pertaining to AC reasoning in \querysmt's proof reconstruction, but this has the disadvantage that the extra facts may lead to proof instability and loss of performance.
\querysmt does not make use of such facts by default, but has an option to enable their inclusion. We discuss the impact of this option in \Cref{sec:component-evaluation}.


\section{Hint Interpretation}
\label{sec:hint-interpretation}

For the most part, interpreting \cvc5's SMT-LIB hints and translating them into Lean expressions is straightforward. At each stage of \leanauto's translation pipeline, \leanauto creates mappings from terms and types in the source language to terms and types in the target language. This is to ensure that when the same term or type appears multiple times in the source problem, each instance of said term or type is translated in the same way. We modified \leanauto's translation procedure to make each of these mappings reversible, allowing \querysmt to translate SMT-LIB symbols and identifiers by simply passing them to the composition of \leanauto's reversed mappings. Although this is sufficient in most cases, there are two complications which merit further discussion.

\paragraph{Non-injectivity.}

\leanauto's mapping from Lean types to SMT-LIB sorts is not injective. \leanauto translates Lean's \texttt{Prop} and \texttt{Bool} types to the same SMT-LIB \texttt{Bool} sort, and as discussed in \Cref{sec:translation}, it also translates Lean's \texttt{Int} and \texttt{Nat} types into the same SMT-LIB \texttt{Int} sort. Consequently, it is possible for \cvc5's hints to contain applications which typecheck according to SMT-LIB's semantics but fail to typecheck when naively translated into Lean. For example, if the original Lean goal contains $n : \texttt{Nat}$ and $x : \texttt{Int}$, \cvc5 may create a hint which involves subtracting $\hat{n}$ from $\hat{x}$, an operation that is unproblematic in SMT-LIB but would fail to typecheck when translated back into Lean.

\querysmt's approach to interpreting such hints consists of defaulting to \texttt{Prop} and \texttt{Int} interpretations of SMT-LIB's \texttt{Bool} and \texttt{Int} sorts, inserting coercions as needed. There is only one circumstance in which expressions must be coerced to \texttt{Bool} or \texttt{Nat}, namely, when supplying an argument to a function that takes \texttt{Bool} or \texttt{Nat} inputs. To give a concrete example, if $f : \texttt{Nat} \rightarrow \texttt{Nat}$, $n : \texttt{Nat}$, and $m : \texttt{Nat}$, then the SMT-LIB hint \texttt{(< $\hat{n}$ ($\hat{f}$ (- $\hat{n}$ $\hat{m}$)))} would be translated to $\texttt{Int.ofNat}\ n < \texttt{Int.ofNat}\ (f\ (\texttt{Int.ofNat}\ n - \texttt{Int.ofNat}\ m)\texttt{.natAbs})$. This is more verbose than the seemingly more natural translation $n < f\ (n - m)$, but a faithful interpretation of \cvc5's hints requires that all built-in mathematical operations occur on integers rather than naturals.\footnote{This is particularly relevant when subtraction is involved, as the semantics of subtraction on the naturals differs from the semantics of subtraction on the integers. For example, the SMT-LIB term \texttt{(+ (- x y) y)} always evaluates to $x$, but the Lean expression $(x - y) + y$ is equal to $\texttt{Nat.max}\ x\ y$ if $x$ and $y$ have type \texttt{Nat}.}

Note that in the previous example, $f$ is not supplied with the (possibly negative) result of subtracting $\texttt{Int.ofNat}\ m$ from $\texttt{Int.ofNat}\ n$. Instead, $f$ is supplied with the absolute value of said result. This coercion is needed to make the Lean expression typecheck, but one might reasonably question whether it compromises the faithfulness of the hint's interpretation. To answer this concern, we observe a subtle consequence of
\Cref{thm:well-formedness}.

\begin{corollary}
    Let $P$ be the SMT-LIB problem obtained by using \textup{\leanauto} to translate a Lean goal, let $f : \textup{\texttt{Nat}} \rightarrow \alpha$ be a Lean function in said goal, and let $\hat{f}$ be the the SMT-LIB translation of $f$ in $P$. For any SMT-LIB formula $F$, if $F$ is entailed by $P$, then $F[(\hat{f} \circ \textup{\texttt{abs}})/\hat{f}]$ is also entailed by $P$.

\end{corollary}

\begin{proofsketch}
    From \Cref{thm:well-formedness}, wherever \texttt{($\hat{f}$ t)} appears in the problem generated by \leanauto, the term \texttt{t} is well-formed. Therefore, any nontrivial assertions about the output of $\hat{f}$ which can be derived from the generated problem are conditioned on $\hat{f}$'s input being nonnegative. Since the output of $\hat{f}$ on negative inputs is unconstrained, any derivable fact about $\hat{f}$ also applies to all functions which agree with $\hat{f}$ on nonnegative inputs. $\hat{f} \circ \texttt{abs}$ agrees with $\hat{f}$ on nonnegative inputs, so any fact that can be derived about $\hat{f}$ also applies to $\hat{f} \circ \texttt{abs}$.
\end{proofsketch}

\paragraph{Non-surjectivity.}

There are two ways SMT-LIB terms without direct Lean equivalents may appear in \cvc5's hints. First, during the course of \cvc5's proof search, \cvc5 may apply Skolemization rules to generate constants which lack direct analogues among the expressions that appear in the input goal. This issue is mitigated by the preprocessing discussed in \Cref{sec:preprocessing}.

The second way SMT-LIB terms without direct Lean equivalents may appear in \cvc5's hints relates to the translation of inductive datatypes. In Lean, a constructor's arguments can be accessed via pattern matching or by invoking a recursor that every inductive type is automatically equipped with. But in SMT-LIB, constructors' arguments are accessed via selector functions whose symbols are given as part of the datatype's declaration. In the special case that the inductive datatype being translated is also a structure, these selector functions can be identified with the projection functions that come with all Lean structures. But when the inductive datatype being translated is not a structure, there is no guarantee that Lean has ready-made analogues for SMT-LIB's selector functions.

In order to interpret hints which refer to selector functions, \querysmt adds fresh functions to the local context along with proofs that they satisfy the property that characterizes SMT-LIB's selector functions. The functions themselves are constructed from the inductive datatype's recursor, and the proofs they are paired with assert that if the function is passed the correct constructor, then the resulting application returns the appropriate argument of said constructor.

\begin{figure}[ht]
\begin{minipage}{0.95\textwidth}\vspace{0pt}
\begin{lstlisting}[language=lean]
example {α : Type} [Inhabited α] (x y : α) : [x] = [y] ↔ x = y := by
  apply @Classical.byContradiction
  intro negGoal
  obtain ⟨_List.cons_sel0, _List.cons_sel0Fact⟩ :
    ∃ (_List.cons_sel0 : List α → α),
      ∀ (arg0 : α) (arg1 : List α),
        _List.cons_sel0 (arg0 :: arg1) = arg0 := by
    apply
      Exists.intro (List.rec (motive := fun (_ : List α) => α)
        default fun (arg0 : α) (arg1 : List α) (_ : α) => arg0)
    intros
    rfl
  duper [negGoal, _List.cons_sel0Fact] []
\end{lstlisting}
\end{minipage}
\caption{A proof script suggested by \querysmt showcasing how Lean analogues for SMT-LIB's selector functions are constructed. Only one of the selector functions for lists is reproduced in the proof script because the other selector function (which retrieves the tail of a nonempty list) is not needed for the proof \duper finds.}
\label{fig:selector-example}
\end{figure}

\Cref{fig:selector-example} provides an example showcasing the construction on a goal with lists. When given the wrong constructor, the function returns \texttt{default}, an arbitrary element of the appropriate type which is only accessible if said type is \texttt{Inhabited}. If typeclass inference cannot prove that the type is \texttt{Inhabited}, then it instead uses \texttt{sorry}, leaving the proof of type inhabitation as a subgoal for the user.


\section{Proof Reconstruction}
\label{sec:proof-reconstruction}

The primary goal of \querysmt's proof reconstruction is not to produce a complete proof term for the given goal. Instead, the goal of \querysmt's proof reconstruction is to suggest a self-contained proof script for the user to examine and potentially modify. All proof scripts suggested by \querysmt consist of:
\begin{enumerate}
    \item A tactic sequence designed to reproduce the effects of \querysmt's preprocessing and Skolemization, described in \Cref{sec:preprocessing}.
    \item A sequence of \texttt{obtain} statements which create functions satisfying the properties of SMT-LIB's selectors. The construction of these functions is described in \Cref{sec:hint-interpretation}.
    \item A sequence of \texttt{have} statements which assert hints output by \cvc5. These \texttt{have} statements are proven with \grind, a built-in Lean tactic.
    \item A final call to \duper \cite{CluneEtAl2024Duper}, a superposition theorem prover intended to reconstruct the logical component of \cvc5's top-level proof.
\end{enumerate}

To minimize the suggested proof script, some sections are omitted if deemed unnecessary. As mentioned in \Cref{sec:preprocessing}, \skolemizeAll is only added to the suggested proof script if Skolemization will change the goal. Additionally, the sequence of \texttt{obtain} and \texttt{have} statements from steps 2 and 3 are minimized by only including those that are necessary for the proof \duper finds in step 4.

\querysmt's ability to perform this minimization, and therefore suggest a usable proof script, depends on \duper successfully finding a proof. Although \duper has been shown to be effective in solving problems previously minimized by other superposition theorem provers \cite{CluneEtAl2024Duper}, its performance degrades significantly when given too many unnecessary or irrelevant premises, and it lacks the theory-specific knowledge leveraged by SMT solvers.

To increase \duper's effectiveness in reconstructing the logical component of \cvc5's proofs, we augment its given clause procedure to implement a variant of the set of support strategy used by Vampire \cite{LPAR-21S:Set_Support_Theory_Reasoning}. This set of support strategy is designed to enable reasoning about theory axioms while mitigating the negative impact of their explosive properties. Facts that are included in \duper's set of support are treated normally, but facts that are excluded from \duper's set of support are only considered when they can be applied to facts in the set of support. The core idea is to limit theory axioms' explosive behavior by only applying them to facts that directly relate to the original goal.

We initially implemented this set of support strategy to exclude \cvc5's hints from the set of support, thinking that \cvc5's hints might behave similarly to more general theory axioms. However, experiments described in \Cref{sec:component-evaluation} revealed that this was actually detrimental to performance. Instead, the set of support strategy is used to exclude a small set of theory lemmas detailing properties of integers and natural numbers that \cvc5's hints aren't expected to capture. The set of theory lemmas used by \querysmt is provided in \Cref{sec:casting-facts}.


\section{Evaluation}

We evaluate \querysmt and existing tools on 9,904 theorems related to integers, natural numbers, and lists taken from Lean's Init, Batteries, and Mathlib \cite{ThemathlibCommunity2020} libraries. Our evaluation focuses on these domains in particular, rather than randomly selected theorems, because we specifically seek to evaluate whether \querysmt benefits from \cvc5's domain-specific knowledge. Int theorems are chosen to test \querysmt's ability to benefit from hints related to SMT-LIB's LIA logic. Nat theorems are chosen to test whether \querysmt can make use of these hints even when it requires encoding Nat goals into Int problems and inferring facts about natural numbers from hints about integers. List theorems are chosen as a proxy for testing \querysmt's ability to make use of SMT solvers' built-in support for reasoning about algebraic datatypes.

\subsection{Methodology}

We perform all experiments on version \texttt{leanprover/lean4:v4.22.0} of Lean. The 9,904 theorems used as benchmark problems are obtained by scraping user-defined theorems from \texttt{Init.Data.X}, \texttt{Batteries.Data.X}, and \texttt{Mathlib.Data.X}, where \texttt{X} is any file prefixed with \texttt{Int}, \texttt{Nat}, or \texttt{List}. A constant is considered a user-defined theorem if it is marked as a theorem, has an explicit declaration in source code, and is not a projection function. All experiments are performed on an Amazon EC2 \texttt{ami-04f167a56786e4b09} instance with 4 virtual CPUs and 16 GiB memory.
Each theorem is given a wall clock timeout of 30 seconds and the default Lean
heartbeat limit of 200,000. The short timeout is used to reflect the expectation
of proof assistant users of having quick results from tactics.

For premise selection, we approximate an ideal premise selector by inspecting the existing proofs of the benchmark theorems and extracting the set of premises $\mathcal{P}$ used to prove them. We also gather the set of constants $\mathcal{C}$ that are not theorems which appear in \texttt{rw} or \texttt{simp} calls of tactic proofs. These constants are used to indicate to the relevant automation that said constant should be unfolded, enabling the automation to invoke definitional equalities not otherwise captured by $\mathcal{P}$. Both \querysmt and the tools we compare with benefit from receiving these constants, so in the experiments, all tools are given $\mathcal{P} \cup \mathcal{C}$ as input.

Our testing script implements the following procedure:

\begin{enumerate}
    \item Identify the (theorem, tool) pair to be tested.
    \item Create a temporary Lean file which imports the benchmark theorem's original file as well as any files needed to run the tool being evaluated.
    \item In the temporary Lean file, define \texttt{alias fakeThm := originalThm}.
    \item Compile the temporary Lean file and extract the \texttt{ConstantInfo} associated with \texttt{fakeThm} along with the environment immediately prior to executing the \texttt{alias} command.
    \item In the environment extracted from the previous step, create a fresh metavariable whose type is determined by the extracted \texttt{ConstantInfo} and attempt to instantiate this metavariable with the tool being tested.\footnote{The tool is considered to have succeeded if the variable is instantiated, regardless of whether the instantiation contains \texttt{sorry}. When \leanauto + \cvc5 is evaluated, the only proof it produces is \texttt{sorry} (indicating that \cvc5 found a proof and \leanauto trusts the result). \querysmt also closes goals with \texttt{sorry} because \querysmt is meant to be replaced with the suggested proof script. Since \querysmt only suggests a proof script if \duper finds a proof that follows from the hints, the suggested script is expected to succeed up to proof reconstruction for the individual hint assertions, for which we have a high success rate.}
\end{enumerate}

We note that the environment in which the tools are run does not perfectly match the original proof's environment. It is infeasible to exactly mimic this environment because among the tools being evaluated, only \grind (which is built into Lean and requires no special imports) would be callable. The differences between the original environments and the environments used in our experiments are largely benign, but in \Cref{sec:results} we discuss one circumstance where the difference in environments is more impactful.

\subsection{Results}
\label{sec:results}

\subsubsection{Tool Comparison}

We compare \querysmt's performance against other tools available in Lean which either interface with external SMT solvers or implement techniques used by SMT solvers. Descriptions of these tools are included in \Cref{fig:methods-description} and their respective performances are shown in \Cref{tbl:benchmark-results}.

\begin{figure}[ht]
\begin{minipage}{0.95\textwidth}
\begin{enumerate}
    \item \leanauto + \cvc5: A tactic which uses \leanauto to translate input problems into the SMT-LIB format and trusts any proofs produced by \cvc5. This serves as a theoretical upper bound for both \querysmt and \leansmt.
    \item \querysmt: The default implementation of \querysmt. \querysmt is considered to succeed if \duper finds a top level proof of the original goals assuming the hints given by \cvc5. This does not necessarily entail that \grind alone is sufficient to prove all the hints \duper depends on. \grind's success rate at proving the hints output by \cvc5 is evaluated separately.
    \item \querysmtminus: A modified implementation of \querysmt in which \duper is not provided the hints output by \cvc5. \querysmtminus retains the preprocessing described in \Cref{sec:preprocessing} and still uses \cvc5's unsat core to minimize the set of premises provided to \duper, but does not translate \cvc5's hints into Lean subgoals or pass the resulting assertions into \duper.
    \item \leansmt: A tactic that interfaces with \cvc5 and performs proof reconstruction via proof replay \cite{Mohamed2025}. At the advice of one of its authors, \leansmt is run with the \texttt{+ mono} option which instructs \leansmt to use \leanauto as a component of its preprocessing.
    \item \grind: A built-in Lean tactic inspired by modern SMT solvers. \grind does not interface with external SMT solvers, but the inspiration for its underlying design and widespread use make it a helpful point of comparison.
\end{enumerate}
\end{minipage}
\caption{Descriptions of SMT-related methods}
\label{fig:methods-description}
\end{figure}

\begin{table}[ht]
\begin{center}
\caption{Benchmark theorems solved by SMT-related methods}
\label{tbl:benchmark-results}
\begin{tabular}{|l|c|c|c|}
\hline
& Int Theorems & Nat Theorems & List Theorems\\
\hline
Total & 2058 & 3270 & 4576 \\
\hline
\leanauto + \cvc5 & 1137 & 1486 & 891\\
\hline
\querysmt & 840 & 892 & 749 \\
\hline
\querysmtminus & 472 & 627 & 708 \\
\hline
\leansmt & 333 & 35 & 445 \\
\hline
\grind & 541 & 812 & - \\
\hline
\end{tabular}
\end{center}
\end{table}

In all categories, \querysmt performs noticeably better with SMT hints than without. On Int and Nat benchmarks, \querysmt only outperforms \grind with these hints. The impact of hints on \querysmt's performance appears to be more significant on Int and Nat benchmarks than on List benchmarks, and not coincidentally, \leanauto + \cvc5 solves a much smaller fraction of List theorems than Int or Nat theorems. From manual inspection of the theorems involved, we suspect that a significant factor contributing to this discrepancy is that in the List category, there is a significant overlap between the set of theorems for which built-in datatype reasoning would be helpful and the set of theorems for which induction is necessary.
\cvc5 is not able to solve problems requiring induction by
default~\cite{Reynolds2015-induction}, or to produce proofs for it, so it cannot
be used by \querysmt in this scenario.
We therefore suspect that a smaller fraction of the List theorems being solved by \leanauto + \cvc5 genuinely require theory reasoning.

On problems relating to Ints and Nats, \querysmt performs best, followed by \grind, followed by \leansmt. We note that \grind has stricter conditions on its input lemmas than \querysmt or \leansmt, and also that \grind benefits from additional hints about how to use its input lemmas (in the form of custom attributes). It may be possible to achieve better performance with \grind either by tailoring the set of provided premises to better suit \grind or by manually providing additional hints about how to use the premises it receives.

\leansmt's performance in both the Int and Nat categories is severely diminished by the fact that \leansmt lacks special support for natural numbers. This says more about temporary limitations resulting from \leansmt's current coverage than the theoretical limit of \leansmt's approach. Still, we note that one of the benefits of our method is relative ease of extensionality. It requires much less effort to add support for parsing hints in a new theory than to add full-fledged proof replay for the same theory.

We omit \grind's performance on List problems in \Cref{tbl:benchmark-results} because it is significantly impacted by our evaluation methodology. When evaluating \grind's performance on List problems with the same script used in the other experiments, \grind succeeds at finding proofs for 1,458 problems. However, upon investigating why \grind performs so much better than the other tools, we discovered that a nontrivial number of problems are solved by \grind accessing lemmas it wouldn't have access to in the original environment. In particular, several theorems tagged with \grind attributes yield benchmark problems that \grind solves by invoking the original theorem. When we modify the evaluation script to test \grind in the original theorem's proof environment\footnote{Note that this test is only feasible because \grind is built into Lean and is therefore accessible in the original proof environment.}, \grind only solves 439 problems. The interpretation of these results depends on whether one views the theorems tagged with \grind attributes as part of \grind's implementation.

\subsubsection{Hint Evaluation}

As noted in \Cref{fig:methods-description}, \querysmt's success criteria depends on \duper deriving a proof of the original goal from the hints output by \cvc5, but does not depend on \grind succeeding at proving all of the subgoals generated by assuming these hints. If \querysmt finds a proof and successfully outputs a proof script in which \grind fails to prove one or more of the generated hints, \querysmt has still done something valuable in reducing the original goal to a smaller subgoal. Still, \querysmt's usefulness is significantly impacted by the frequency with which \cvc5's hints can be proven automatically, so we perform an additional evaluation to test how frequently \cvc5 produces hints that \grind can't solve. This evaluation includes not just hints that appear in \querysmt's suggestions, but all hints output by \cvc5 regardless of whether \duper succeeds in its proof search, and regardless of whether they would actually appear in the final proof script suggestion.

Of the 9,904 problems tested, 499 produce a set of hints that \grind fails to certify, approximately 5\% of the total. 57 of these failures come from Int problems, 168 of these failures come from Nat problems, and 274 of these failures come from List problems. From manual inspection, we know that many of these failures are false negatives owed to \cvc5 producing large sets of hints that can be solved individually by \grind but collectively cause \grind to time out\footnote{As in other experiments, \grind gets 200,000 heartbeats and 30 seconds per theorem.}. We still register such cases as failures because it is difficult to distinguish this behavior from tests in which \grind genuinely times out on a single hint. Not all failures are false negatives, but anecdotally, the hints \grind genuinely fails on tend to be easy to discharge manually using some combination of \aesop \cite{Limperg2023} and \duper.

We also measure the number of hints generated for problems where \querysmt succeeds in suggesting a proof script. Proof scripts with many hints are harder to read, modify, and maintain than proof scripts with just a few hints, so it is preferred for \querysmt to include as few hints as possible in the final suggestion.

The average number of hints \cvc5 generates on Int, Nat, and List problems that \querysmt solves is 3.1, 4.6, and 4.6 respectively. After filtering out hints that are not needed for the final proof, \querysmt's final suggestion only includes 0.8, 0.7, and 0.2 hints on average for Int, Nat, and List problems respectively. These averages are brought down by the problems that \duper can solve without any hints, but even after filtering out all problems in which \querysmt produces 0 hints, the average number of hints produced is only 1.5, 1.7, and 1.2 for Int, Nat, and List problems respectively. This does not mitigate all readability concerns, as individual hints can still be unnecessarily verbose, but it does show that \querysmt tends to produce suggestions of manageable size.



\section{Conclusion}

We explored a new hint-based approach to leveraging SMT solvers for ITP automation. We implemented this approach in the Lean proof assistant to create \querysmt, a tactic that translates Lean goals to SMT-LIB, extracts the preprocessing and theory reasoning used by \cvc5 to solve the translated problem, and uses that information to produce a self-contained proof script for the original goal which does not depend on \cvc5. We evaluated \querysmt on problems related to its supported theories, showing that \querysmt compares favorably to existing SMT-related Lean automation and that the hints extracted from \cvc5 produce a clear improvement in the underlying proof automation.

We see several possible directions for future work. One possibility is to implement our approach in other proof assistants or SMT solvers to see whether it can be applied to more than just Lean and \cvc5. Another is to add support for more SMT theories to see how our approach generalizes beyond hints relating to integers and algebraic datatypes.
A third possibility is to explore ways to gather even more information from
\cvc5's proofs. For example, collecting instances generated for quantified
formulas could lead to \duper finding proofs more quickly, as was
recently done for \metis in Isabelle's Sledgehammer~\cite{Bartl2025}.
We also expect that \querysmt's reconstruction success rate could be
increased with better instrumentation for tracking how \cvc5 normalizes AC
operators.


\subsubsection*{Acknowledgements}
We thank Hanna Lachnitt for discussing the translation described in \Cref{sec:naturals-translation} and for bringing the \texttt{Trakt} paper \cite{Blot2023} to our attention. We also thank the anonymous reviewers for their feedback on this paper. This work was partially supported by funding from AFRL and DARPA under Agreement FA8750-24-9-1000.

\subsubsection*{Data-Availability Statement} The source code for \querysmt is available at \url{https://github.com/JOSHCLUNE/QuerySMT}. An artifact with the code necessary to replicate our experiments is available at \url{https://zenodo.org/records/18190143}.

\bibliographystyle{splncs04.bst}
\bibliography{main.bib}

@string{itp  = "Interactive Theorem Proving (ITP)"}

@string{cade   = "Proc. Conference on Automated Deduction (CADE)"}

@string{tacas  = "Tools and Algorithms for Construction and Analysis of
                 Systems (TACAS)"}

@string{tacas1  = "Tools and Algorithms for Construction and Analysis of
                 Systems (TACAS), Part I"}

@string{fmcad  = "Formal Methods In Computer-Aided Design (FMCAD)"}

@string{ijcar  = "International Joint Conference on Automated Reasoning
                 (IJCAR)"}

@inbook{qian2025lean-auto,
  title = {Lean-Auto: An Interface Between Lean 4 and Automated Theorem Provers},
  ISBN = {9783031986826},
  ISSN = {1611-3349},
  DOI = {10.1007/978-3-031-98682-6_10},
  booktitle = {Computer Aided Verification},
  publisher = {Springer Nature Switzerland},
  author = {Qian,  Yicheng and Clune,  Joshua and Barrett,  Clark and Avigad,  Jeremy},
  year = {2025},
  pages = {175–196}
}

@misc{Theorem_Proving_in_Lean4,
    title = {Theorem Proving in {Lean 4}},
    url = {https://leanprover.github.io/theorem\_proving\_in\_lean4/},
    author = {Avigad, Jeremy and de Moura, Leonardo and Kong, Soonho and Ullrich, Sebastian}
}

@techreport{BarFT-RR-17,
  author = {Clark Barrett and Pascal Fontaine and Cesare Tinelli},
  institution = {Department of Computer Science, The University of Iowa},
  note = {Available at {\tt www.SMT-LIB.org}},
  title = {{The SMT-LIB Standard: Version 2.6}},
  year = 2017
}

@techreport{BarFT-RR-25,
  author = {Clark Barrett and Pascal Fontaine and Cesare Tinelli},
  institution = {Department of Computer Science, The University of Iowa},
  note = {Available at {\tt www.SMT-LIB.org}},
  title = {{The SMT-LIB Standard: Version 2.7}},
  year = 2025
}

@InProceedings{CluneEtAl2024Duper,
  author =  {Clune, Joshua and Qian, Yicheng and Bentkamp, Alexander and Avigad, Jeremy},
  title = {{Duper: A Proof-Producing Superposition Theorem Prover for Dependent Type Theory}},
  booktitle = {Interactive Theorem Proving (ITP)},
  pages = {10:1--10:20},
  year =  {2024},
  volume =  {309},
  editor =  {Bertot, Yves and Kutsia, Temur and Norrish, Michael},
  publisher = {Schloss Dagstuhl -- Leibniz-Zentrum f{\"u}r Informatik},
  address = {Dagstuhl, Germany},
  doi =   {10.4230/LIPIcs.ITP.2024.10},
}

@inbook{Mohamed2025,
  title = {lean-smt: An SMT Tactic for Discharging Proof Goals in Lean},
  ISBN = {9783031986826},
  ISSN = {1611-3349},
  url = {http://dx.doi.org/10.1007/978-3-031-98682-6\_11},
  DOI = {10.1007/978-3-031-98682-6_11},
  booktitle = {Computer Aided Verification},
  publisher = {Springer Nature Switzerland},
  author = {Mohamed,  Abdalrhman and Mascarenhas,  Tomaz and Khan,  Harun and Barbosa,  Haniel and Reynolds,  Andrew and Qian,  Yicheng and Tinelli,  Cesare and Barrett,  Clark},
  year = {2025},
  pages = {197–212}
}

@inbook{Armand2011,
  title = {A Modular Integration of SAT/SMT Solvers to Coq through Proof Witnesses},
  ISBN = {9783642253799},
  ISSN = {1611-3349},
  DOI = {10.1007/978-3-642-25379-9_12},
  booktitle = {Certified Programs and Proofs},
  publisher = {Springer Berlin Heidelberg},
  author = {Armand,  Michael and Faure,  Germain and Grégoire,  Benjamin and Keller,  Chantal and Théry,  Laurent and Werner,  Benjamin},
  year = {2011},
  pages = {135–150}
}

@article{MengP08,
  author       = {Jia Meng and
                  Lawrence C. Paulson},
  title        = {Translating Higher-Order Clauses to First-Order Clauses},
  journal      = {J. Autom. Reason.},
  volume       = {40},
  number       = {1},
  pages        = {35--60},
  year         = {2008},
  OPTurl          = {https://doi.org/10.1007/s10817-007-9085-y},
  doi          = {10.1007/S10817-007-9085-Y},
}

@article{MengP09,
  author       = {Jia Meng and
                  Lawrence C. Paulson},
  title        = {Lightweight relevance filtering for machine-generated resolution problems},
  journal      = {J. Appl. Log.},
  volume       = {7},
  number       = {1},
  pages        = {41--57},
  year         = {2009},
  OPTurl          = {https://doi.org/10.1016/j.jal.2007.07.004},
  doi          = {10.1016/J.JAL.2007.07.004},
}

@inproceedings{PaulsonS07,
  author       = {Lawrence C. Paulson and
                  Kong Woei Susanto},
  editor       = {Klaus Schneider and
                  Jens Brandt},
  title        = {Source-Level Proof Reconstruction for Interactive Theorem Proving},
  booktitle    = {Theorem Proving in Higher Order Logics, 20th International Conference,
                  TPHOLs 2007, Kaiserslautern, Germany, September 10-13, 2007, Proceedings},
  series       = {Lecture Notes in Computer Science},
  volume       = {4732},
  pages        = {232--245},
  publisher    = {Springer},
  year         = {2007},
  OPTurl          = {https://doi.org/10.1007/978-3-540-74591-4\_18},
  doi          = {10.1007/978-3-540-74591-4_18},
}

@inproceedings{PaulsonB10,
  author       = {Lawrence C. Paulson and
                  Jasmin Christian Blanchette},
  editor       = {Geoff Sutcliffe and
                  Stephan Schulz and
                  Eugenia Ternovska},
  title        = {Three years of experience with Sledgehammer, a Practical Link Between
                  Automatic and Interactive Theorem Provers},
  booktitle    = {The 8th International Workshop on the Implementation of Logics, {IWIL}
                  2010, Yogyakarta, Indonesia, October 9, 2011},
  series       = {EPiC Series in Computing},
  volume       = {2},
  pages        = {1--11},
  publisher    = {EasyChair},
  year         = {2010},
  OPTurl          = {https://doi.org/10.29007/36dt},
  doi          = {10.29007/36DT},
}

@article{Czajka2018,
  title = {Hammer for Coq: Automation for Dependent Type Theory},
  volume = {61},
  ISSN = {1573-0670},
  DOI = {10.1007/s10817-018-9458-4},
  number = {1–4},
  journal = {Journal of Automated Reasoning},
  publisher = {Springer Science and Business Media LLC},
  author = {Czajka,  Lukasz and Kaliszyk,  Cezary},
  year = {2018},
  month = feb,
  pages = {423–453}
}

@inproceedings{Blot2023,
  series = {CPP ’23},
  title = {Compositional Pre-processing for Automated Reasoning in Dependent Type Theory},
  DOI = {10.1145/3573105.3575676},
  booktitle = {Proceedings of the 12th ACM SIGPLAN International Conference on Certified Programs and Proofs},
  publisher = {ACM},
  author = {Blot,  Valentin and Cousineau,  Denis and Crance,  Enzo and de Prisque,  Louise Dubois and Keller,  Chantal and Mahboubi,  Assia and Vial,  Pierre},
  year = {2023},
  month = jan,
  pages = {63–77},
  collection = {CPP ’23}
}

@inproceedings{ThemathlibCommunity2020,
  editor       = {Jasmin Blanchette and
                  Catalin Hritcu},
  author       = {The mathlib Community},
  title        = {The {Lean} mathematical library},
  booktitle    = {Proceedings of the 9th {ACM} {SIGPLAN} International Conference on
                  Certified Programs and Proofs, {CPP} 2020, New Orleans, LA, USA, January
                  20-21, 2020},
  pages        = {367--381},
  publisher    = {{ACM}},
  year         = {2020},
  OPTurl       = {https://doi.org/10.1145/3372885.3373824},
  doi          = {10.1145/3372885.3373824},
  bibsource    = {dblp computer science bibliography, https://dblp.org}
}

@article{Blanchette2016-qed,
  author    = {Jasmin Christian Blanchette and
               Cezary Kaliszyk and
               Lawrence C. Paulson and
               Josef Urban},
  title     = {Hammering towards {QED}},
  journal   = {J. Formalized Reasoning},
  volume    = {9},
  number    = {1},
  pages     = {101--148},
  year      = {2016}
}

@inproceedings{DBLP:conf/cade/BlanchetteBP11,
  author       = {Jasmin Christian Blanchette and
                  Sascha B{\"{o}}hme and
                  Lawrence C. Paulson},
  editor       = {Nikolaj S. Bj{\o}rner and
                  Viorica Sofronie{-}Stokkermans},
  title        = {Extending Sledgehammer with {SMT} Solvers},
  booktitle    = {Automated Deduction - {CADE-23} - 23rd International Conference on
                  Automated Deduction, Wroclaw, Poland, July 31 - August 5, 2011. Proceedings},
  series       = {Lecture Notes in Computer Science},
  volume       = {6803},
  pages        = {116--130},
  publisher    = {Springer},
  year         = {2011},
  url          = {https://doi.org/10.1007/978-3-642-22438-6\_11},
  doi          = {10.1007/978-3-642-22438-6\_11},
  bibsource    = {dblp computer science bibliography, https://dblp.org}
}

@inproceedings{DBLP:conf/itp/NormanA25,
  author       = {Chase Norman and
                  Jeremy Avigad},
  editor       = {Yannick Forster and
                  Chantal Keller},
  title        = {Canonical for Automated Theorem Proving in Lean},
  booktitle    = {16th International Conference on Interactive Theorem Proving, {ITP}
                  2025, September 28 to October 1, 2025, Reykjavik, Iceland},
  series       = {LIPIcs},
  volume       = {352},
  pages        = {14:1--14:20},
  publisher    = {Schloss Dagstuhl - Leibniz-Zentrum f{\"{u}}r Informatik},
  year         = {2025},
  url          = {https://doi.org/10.4230/LIPIcs.ITP.2025.14},
  doi          = {10.4230/LIPICS.ITP.2025.14},
  bibsource    = {dblp computer science bibliography, https://dblp.org}
}

@incollection{Barrett2021,
  author    = {Clark W. Barrett and
               Roberto Sebastiani and
               Sanjit A. Seshia and
               Cesare Tinelli},
  editor    = {Armin Biere and
               Marijn Heule and
               Hans van Maaren and
               Toby Walsh},
  title     = {Satisfiability Modulo Theories},
  booktitle = {Handbook of Satisfiability - Second Edition},
  series    = {Frontiers in Artificial Intelligence and Applications},
  volume    = {336},
  pages     = {1267--1329},
  publisher = {{IOS} Press},
  year      = {2021},
  url       = {https://doi.org/10.3233/FAIA201017},
  doi       = {10.3233/FAIA201017},
  bibsource = {dblp computer science bibliography, https://dblp.org}
}

@inproceedings{Schurr2021,
  author    = {Hans{-}J{\"{o}}rg Schurr and
               Mathias Fleury and
               Martin Desharnais},
  editor    = {Andr{\'{e}} Platzer and
               Geoff Sutcliffe},
  title     = {Reliable Reconstruction of Fine-grained Proofs in a Proof Assistant},
  booktitle = cade,
  series    = {Lecture Notes in Computer Science},
  volume    = {12699},
  pages     = {450--467},
  publisher = {Springer},
  year      = {2021},
  url       = {https://doi.org/10.1007/978-3-030-79876-5\_26},
  doi       = {10.1007/978-3-030-79876-5\_26},
  bibsource = {dblp computer science bibliography, https://dblp.org}
}

@inproceedings{Lachnitt2025,
  author       = {Hanna Lachnitt and
                  Mathias Fleury and
                  Haniel Barbosa and
                  Jibiana Jakpor and
                  Bruno Andreotti and
                  Andrew Reynolds and
                  Hans{-}J{\"{o}}rg Schurr and
                  Clark W. Barrett and
                  Cesare Tinelli},
  editor       = {Yannick Forster and
                  Chantal Keller},
  title        = {Improving the {SMT} Proof Reconstruction Pipeline in Isabelle/HOL},
  booktitle    = itp,
  series       = {LIPIcs},
  volume       = {352},
  pages        = {26:1--26:22},
  publisher    = {Schloss Dagstuhl - Leibniz-Zentrum f{\"{u}}r Informatik},
  year         = {2025},
  url          = {https://doi.org/10.4230/LIPIcs.ITP.2025.26},
  doi          = {10.4230/LIPICS.ITP.2025.26},
  bibsource    = {dblp computer science bibliography, https://dblp.org}
}

@inproceedings{Bouton2009,
  author    = {Thomas Bouton and
               Diego Caminha B. de Oliveira and
               David D{\'{e}}harbe and
               Pascal Fontaine},
  title     = {{veriT}: {A}n {O}pen, {T}rustable and {E}fficient {SMT-S}olver},
  booktitle = cade,
  pages     = {151--156},
  year      = {2009},
  url       = {http://dx.doi.org/10.1007/978-3-642-02959-2_12},
  doi       = {10.1007/978-3-642-02959-2_12},
  editor    = {Renate A. Schmidt},
  series    = {Lecture Notes in Computer Science},
  volume    = {5663},
  publisher = {Springer},
}

@inproceedings{Barbosa2022-cvc5,
  author    = {Haniel Barbosa and
               Clark W. Barrett and
               Martin Brain and
               Gereon Kremer and
               Hanna Lachnitt and
               Makai Mann and
               Abdalrhman Mohamed and
               Mudathir Mohamed and
               Aina Niemetz and
               Andres N{\"{o}}tzli and
               Alex Ozdemir and
               Mathias Preiner and
               Andrew Reynolds and
               Ying Sheng and
               Cesare Tinelli and
               Yoni Zohar},
  editor    = {Dana Fisman and
               Grigore Rosu},
  title     = {cvc5: {A} Versatile and Industrial-Strength {SMT} Solver},
  booktitle = tacas1,
  series    = {Lecture Notes in Computer Science},
  volume    = {13243},
  pages     = {415--442},
  publisher = {Springer},
  year      = {2022},
  url       = {https://doi.org/10.1007/978-3-030-99524-9\_24},
  doi       = {10.1007/978-3-030-99524-9\_24},
  bibsource = {dblp computer science bibliography, https://dblp.org}
}

@inproceedings{deMoura2008-z3,
  author    = {Leonardo Mendon{\c{c}}a de Moura and
               Nikolaj Bj{\o}rner},
  editor    = {C. R. Ramakrishnan and
               Jakob Rehof},
  title     = {{Z3:} An Efficient {SMT} Solver},
  booktitle = tacas,
  series    = {Lecture Notes in Computer Science},
  volume    = {4963},
  pages     = {337--340},
  publisher = {Springer},
  year      = {2008},
  url       = {https://doi.org/10.1007/978-3-540-78800-3\_24},
  doi       = {10.1007/978-3-540-78800-3\_24},
  bibsource = {dblp computer science bibliography, https://dblp.org}
}

@inproceedings{LPAR-21S:Set_Support_Theory_Reasoning,
  author    = {Giles Reger and Martin Suda},
  title     = {Set of Support for Theory Reasoning},
  booktitle = {IWIL Workshop and LPAR Short Presentations},
  editor    = {Thomas Eiter and David Sands and Geoff Sutcliffe and Andrei Voronkov},
  series    = {Kalpa Publications in Computing},
  volume    = {1},
  publisher = {EasyChair},
  bibsource = {EasyChair, https://easychair.org},
  issn      = {2515-1762},
  url       = {/publications/paper/4Sd},
  doi       = {10.29007/ndjg},
  pages     = {124-134},
  year      = {2017}
}

@inproceedings{Zhou2023,
  author       = {Yi Zhou and
                  Jay Bosamiya and
                  Yoshiki Takashima and
                  Jessica Li and
                  Marijn Heule and
                  Bryan Parno},
  editor       = {Alexander Nadel and
                  Kristin Yvonne Rozier},
  title        = {Mariposa: Measuring {SMT} Instability in Automated Program Verification},
  booktitle    = fmcad,
  pages        = {178--188},
  publisher    = {{IEEE}},
  year         = {2023},
  url          = {https://doi.org/10.34727/2023/isbn.978-3-85448-060-0\_26},
  doi          = {10.34727/2023/ISBN.978-3-85448-060-0\_26},
  bibsource    = {dblp computer science bibliography, https://dblp.org}
}

@inproceedings{Bartl2025,
  author       = {Lukas Bartl and
                  Jasmin Blanchette and
                  Tobias Nipkow},
  editor       = {Clark W. Barrett and
                  Uwe Waldmann},
  title        = {Exploiting Instantiations from Paramodulation Proofs in Isabelle/HOL},
  booktitle    = cade,
  series       = {Lecture Notes in Computer Science},
  volume       = {15943},
  pages        = {573--593},
  publisher    = {Springer},
  year         = {2025},
  url          = {https://doi.org/10.1007/978-3-031-99984-0\_30},
  doi          = {10.1007/978-3-031-99984-0\_30},
  bibsource    = {dblp computer science bibliography, https://dblp.org}
}

@inproceedings{Limperg2023,
  author       = {Jannis Limperg and
                  Asta Halkj{\ae}r From},
  editor       = {Robbert Krebbers and
                  Dmitriy Traytel and
                  Brigitte Pientka and
                  Steve Zdancewic},
  title        = {Aesop: White-Box Best-First Proof Search for Lean},
  booktitle    = {Proceedings of the 12th {ACM} {SIGPLAN} International Conference on
                  Certified Programs and Proofs, {CPP} 2023, Boston, MA, USA, January
                  16-17, 2023},
  pages        = {253--266},
  publisher    = {{ACM}},
  year         = {2023},
  OPTurl          = {https://doi.org/10.1145/3573105.3575671},
  doi          = {10.1145/3573105.3575671},
  bibsource    = {dblp computer science bibliography, https://dblp.org}
}

@inproceedings{Kovcs2013,
  author       = {Laura Kov{\'{a}}cs and
                  Andrei Voronkov},
  editor       = {Natasha Sharygina and
                  Helmut Veith},
  title        = {First-Order Theorem Proving and Vampire},
  booktitle    = {Computer Aided Verification - 25th International Conference, {CAV}
                  2013, Saint Petersburg, Russia, July 13-19, 2013. Proceedings},
  series       = {Lecture Notes in Computer Science},
  volume       = {8044},
  pages        = {1--35},
  publisher    = {Springer},
  year         = {2013},
  OPTurl          = {https://doi.org/10.1007/978-3-642-39799-8\_1},
  doi          = {10.1007/978-3-642-39799-8\_1},
  bibsource    = {dblp computer science bibliography, https://dblp.org}
}

@inbook{Vukmirovi2021,
  title = {Making Higher-Order Superposition Work},
  ISBN = {9783030798765},
  ISSN = {1611-3349},
  OPTurl = {http://dx.doi.org/10.1007/978-3-030-79876-5_24},
  DOI = {10.1007/978-3-030-79876-5_24},
  booktitle = {Lecture Notes in Computer Science},
  publisher = {Springer International Publishing},
  author = {Vukmirović,  Petar and Bentkamp,  Alexander and Blanchette,  Jasmin and Cruanes,  Simon and Nummelin,  Visa and Tourret,  Sophie},
  year = {2021},
  pages = {415–432}
}

@inproceedings{hurd2003d,
  author = {Joe Hurd},
  title = {First-Order Proof Tactics in Higher-Order Logic Theorem Provers},
  pages = {56--68},
  booktitle = {Design and Application of Strategies/Tactics in Higher Order Logics (STRATA 2003)},
  url = {http://www.gilith.com/papers},
  year = 2003,
}

@inbook{Vukmirovi2023,
  title = {Extending a High-Performance Prover to Higher-Order Logic},
  ISBN = {9783031308208},
  ISSN = {1611-3349},
  url = {http://dx.doi.org/10.1007/978-3-031-30820-8_10},
  DOI = {10.1007/978-3-031-30820-8_10},
  booktitle = {Tools and Algorithms for the Construction and Analysis of Systems},
  publisher = {Springer Nature Switzerland},
  author = {Vukmirović,  Petar and Blanchette,  Jasmin and Schulz,  Stephan},
  year = {2023},
  pages = {111–129}
}

@article{Bving2025,
  title = {Interactive Bitvector Reasoning using Verified Bit-Blasting},
  volume = {9},
  ISSN = {2475-1421},
  url = {http://dx.doi.org/10.1145/3763167},
  DOI = {10.1145/3763167},
  number = {OOPSLA2},
  journal = {Proceedings of the ACM on Programming Languages},
  publisher = {Association for Computing Machinery (ACM)},
  author = {B\"{o}ving,  Henrik and Bhat,  Siddharth and Cicolini,  Luisa and Keizer,  Alex and Frenot,  Léon and Mohamed,  Abdalrhman and Stefanesco,  Léo and Khan,  Harun and Clune,  Joshua and Barrett,  Clark and Grosser,  Tobias},
  year = {2025},
  month = oct,
  pages = {3259–3285}
}

@inproceedings{Barbosa2022,
  author    = {Haniel Barbosa and
               Andrew Reynolds and
               Gereon Kremer and
               Hanna Lachnitt and
               Aina Niemetz and
               Andres N{\"{o}}tzli and
               Alex Ozdemir and
               Mathias Preiner and
               Arjun Viswanathan and
               Scott Viteri and
               Yoni Zohar and
               Cesare Tinelli and
               Clark W. Barrett},
  editor    = {Jasmin Blanchette and
               Laura Kov{\'{a}}cs and
               Dirk Pattinson},
  title     = {Flexible Proof Production in an Industrial-Strength {SMT} Solver},
  booktitle = ijcar,
  series    = {Lecture Notes in Computer Science},
  volume    = {13385},
  pages     = {15--35},
  publisher = {Springer},
  year      = {2022},
  url       = {https://doi.org/10.1007/978-3-031-10769-6\_3},
  doi       = {10.1007/978-3-031-10769-6\_3},
  bibsource = {dblp computer science bibliography, https://dblp.org}
}

@incollection{Reynolds2015-induction,
year={2015},
isbn={978-3-662-46080-1},
booktitle={Verification, Model Checking, and Abstract Interpretation},
volume={8931},
series={Lecture Notes in Computer Science},
editor={D’Souza, Deepak and Lal, Akash and Larsen, KimGuldstrand},
doi={10.1007/978-3-662-46081-8_5},
title={Induction for SMT Solvers},
url={http://dx.doi.org/10.1007/978-3-662-46081-8_5},
publisher={Springer Berlin Heidelberg},
author={Reynolds, Andrew and Kuncak, Viktor},
pages={80-98},
language={English}
}

\appendix
\section{High-Level Proof Sketch of Theorem 1}\label{pf:well-formedness}

\textbf{Theorem 1.}
\textit{All terms except datatype selectors which appear in an SMT-LIB problem generated by our procedure are provably well-formed.}

\begin{proofsketch}
    Let $t : \hat{\alpha}$ be some term which appears in the generated SMT-LIB problem. We proceed by induction on $t$:
    \begin{itemize}
        \item If $t$ is a special constant (i.e. a numeral, decimal, hexadecimal, binary, or string), then either $t$ is interpreted to be of sort \texttt{Int} and is nonnegative, or $t$ is interpreted to be of a sort such that $\texttt{wf}_\alpha\ t = \texttt{True}$.
        \item If $t$ is an identifier corresponding to a declared function or constant, then when said function or constant was declared, an assertion was added to guarantee that $t$ is well-formed.
        \item If $t$ is a symbol corresponding to a bound variable introduced by \texttt{forall} or \texttt{exists}, then when the original Lean quantifier was translated, $\texttt{wf}_\alpha\ t$ was added to the translated body (either as a hypothesis for a universal quantifier's body or in conjunction with an existential quantifier's body).
        \item If $t$ is a symbol corresponding to a datatype's constructor, then $\alpha$ must have the form $\beta_1 \rightarrow \ldots \rightarrow \beta_n \rightarrow \gamma$ where $\beta_1, \ldots \beta_n$ are the types of the Lean constructor's inputs and $\gamma$ is either an inductive datatype or a structure.
        From the definition of well-formedness on function types, $\texttt{wf}_\alpha\ t$ is equivalent to \texttt{(=> (wf$_{\beta_1}$ y1) $\ldots$ (wf$_{\beta_n}$ yn) (wf$_{\gamma}$ (t y1 $\ldots$ yn)))} where \texttt{y1 $\ldots$ yn} are arbitrary terms of sorts $\hat{\beta_1}, \ldots \hat{\beta_n}$.
        \begin{itemize}
            \item If $\gamma$ is a structure, then \texttt{wf$_\gamma$ ($t$ y1 $\ldots$ yn)} asserts that all projections of \texttt{($t$ y1 $\ldots$ yn)} are well-formed. This follows from the implication's hypotheses \texttt{(wf$_{\beta_1}$ y1) $\ldots$ (wf$_{\beta_n}$ yn)}, so $\texttt{wf}_\alpha\ t$ holds.
            \item If $\gamma$ is an inductive datatype, then \texttt{wf$_\gamma$ ($t$ y1 $\ldots$ yn)} asserts that for each of $\gamma$'s constructors $c_i$, if \texttt{($t$ y1 $\ldots$ yn)} satisfies \texttt{(\_ is $\hat{c_i}$)}, then for each selector $s_{i, j}$ of $\hat{c_i}$, \texttt{($s_{i, j}$ ($t$ y1 $\ldots$ yn))} must be well-formed. Since \texttt{($t$ y1 $\ldots$ yn)} satisfies exactly one tester, namely \texttt{(\_ is $t$)}, it follows that \texttt{wf$_\gamma$ ($t$ y1 $\ldots$ yn)} asserts that applying any of $t$'s selectors to \texttt{($t$ y1 $\ldots$ yn)} yields a well-formed term. This follows from the implication's hypotheses \texttt{(wf$_{\beta_1}$ y1) $\ldots$ (wf$_{\beta_n}$ yn)}, so $\texttt{wf}_\alpha\ t$ holds.
        \end{itemize}
        \item If $t$ is a symbol corresponding to a datatype's selector, then \Cref{thm:well-formedness} does not require that $t$ be well-formed.
        \item If $t$ is the \texttt{ite} function from SMT-LIB's \texttt{Core} theory, then $\alpha$ must have the form $\texttt{Prop} \rightarrow \beta \rightarrow \beta \rightarrow \beta$ for some type $\beta$. From the definition of well-formedness on function types, \texttt{wf$_\alpha$ t} is equivalent to \texttt{(=>} \texttt{(wf$_\texttt{Prop}$ y1)} \texttt{(wf$_\beta$ y2)} \texttt{(wf$_\beta$ y3)} \texttt{(wf$_\beta$ (ite y1 y2 y3)))} where \texttt{y1} is an arbitrary term of sort \texttt{Bool} and \texttt{y2} and \texttt{y3} are arbitrary terms of sort $\hat{\beta}$. \texttt{(ite y1 y2 y3)} evaluates to \texttt{y2} or \texttt{y3} depending on \texttt{y1}, and the hypotheses of the implication give \texttt{(wf$_\beta$ y2)} and \texttt{(wf$_\beta$ y3)}, so the conclusion \texttt{(wf$_\beta$ (ite y1 y2 y3))} follows from the implication's hypotheses. This entails \texttt{wf$_\alpha$ $t$} as desired.
        \item If $t \in \{\texttt{+}, \texttt{*}, \texttt{div}, \texttt{mod}, \texttt{abs}\}$, then the semantics of SMT-LIB's \texttt{Int} theory guarantee that $t$ preserves nonnegativity and is therefore well-formed.
        \item If $t$ is the ``\texttt{-}'' symbol and corresponds to negation, then $\alpha = \texttt{Int} \rightarrow \texttt{Int}$ so \texttt{wf$_\alpha$ $t$} is equivalent to \texttt{True}.
        \item If $t$ is the ``\texttt{-}'' symbol and corresponds to subtraction, then $\alpha = \texttt{Int} \rightarrow \texttt{Int} \rightarrow \texttt{Int}$, entailing that \texttt{wf$_\alpha$ $t$} is equivalent to \texttt{True}. Note that $\alpha \ne \texttt{Nat} \rightarrow \texttt{Nat} \rightarrow \texttt{Nat}$ even though subtraction on the naturals exists in Lean because \leanauto does not translate Lean subtraction on the naturals directly into SMT-LIB subtraction on the integers. \leanauto translates $(e_1 : \texttt{Nat}) - (e_2 : \texttt{Nat})$ into the term \texttt{(ite (>= $\hat{e_1}$ $\hat{e_2}$) (- $\hat{e_1}$ $\hat{e_2}$) 0)}, and in this term, \texttt{(- $\hat{e_1}$ $\hat{e_2}$)} can be understood as corresponding to $(\uparrow e_1 : \texttt{Int}) - (\uparrow e_2 : \texttt{Int})$, meaning that even when subtraction on the naturals is being translated to SMT-LIB, the ``\texttt{-}'' symbol that appears in the resulting translation still corresponds to a function of type $\alpha = \texttt{Int} \rightarrow \texttt{Int} \rightarrow \texttt{Int}$.
        \item If $t$ is a symbol corresponding to a built-in function from SMT-LIB's \texttt{Core} or \texttt{Int} theories and $t \notin \{\texttt{-}, \texttt{+}, \texttt{*}, \texttt{div}, \texttt{mod}, \texttt{abs}, \texttt{ite}\}$, then $t$ is a function whose output sort is \texttt{Bool}. This entails that \texttt{wf$_\alpha$ $t$} is equivalent to \texttt{True}.
        \item If $t$ is an application of of the form \texttt{($t_1$ $t_2$)}, then the Lean term corresponding to $t_1$ has type $\beta \rightarrow \alpha$ and the Lean term corresponding to $t_2$ has type $\beta$. By the inductive hypothesis, $t_2$ is well-formed and $t_1$ is either well-formed or a selector function.
        \begin{itemize}
            \item If $t_1$ is well-formed, then from the definition of well-formedness on functions, \texttt{wf$_{\beta \rightarrow \alpha}$ $t_1$ $=$ (forall ((y $\hat{\beta}$)) (=> (wf$_\beta$ y) (wf$_\alpha$ ($t_1$ y))))}.\\
            From this and \texttt{wf$_\beta$ $t_2$}, it follows that \texttt{wf$_\alpha$ ($t_1$ $t_2$)} as desired.
            \item If $t_1$ is a selector function, then $\beta$ is a structure or inductive datatype and $t_1$ has some associated constructor $\hat{c}$. From the definition of well-formedness on structures and inductive datatypes, \texttt{wf$_\beta$ $t_2$} entails \texttt{(=> ($t_2$ is $\hat{c}$) (wf$_\alpha$ ($t_1$ $t_2$)))}. \leanauto's translation procedure guarantees that if \texttt{($t_1$ $t_2$)} appears in the generated SMT-LIB problem, then $t_2$ satisfies the tester for $t_1$'s constructor, meaning \texttt{($t_2$ is $\hat{c}$)} holds. From this and the implication entailed by \texttt{wf$_\beta$ $t_2$}, it follows that \texttt{wf$_\alpha$ ($t_1$ $t_2$)}.
        \end{itemize}
        \item If $t$ is a \texttt{forall} or \texttt{exists} binder, then $\alpha = \texttt{Prop}$ and \texttt{wf$_\alpha$ $t$ $=$ True}.
        \item If $t$ is a \texttt{match} binder, then $t$ evaluates to one of the top-level terms among its match cases. None of these top-level terms can be unapplied selectors, so by the inductive hypothesis, they are all well-formed. Since $t$ evaluates to a well-formed term, $t$ itself is well-formed.
        \item If $t$ is a wrapper expression of the form \texttt{(! $t'$ \ldots)}, then by the inductive hypothesis, $t'$ is well-formed or a selector function. \leanauto's translation procedure only produces wrapper expressions of the form \texttt{(! $t'$ \ldots)} to indicate named formulas, so $t'$ cannot be a selector function. Therefore, $t'$ is well-formed. Since $t$ is semantically equivalent to $t'$, $t$ is also well-formed.
        \item The only remaining term forms to consider are \texttt{let} binders and \texttt{lambda} binders. $t$ is guaranteed to be neither of these because \leanauto's translation procedure never produces them. \texttt{let} binders are never produced because it is never necessary to. \texttt{lambda} binders are never produced because they are part of a recent extension to SMT-LIB's logic which \leanauto does not specifically target.
    \end{itemize}
\end{proofsketch}


\section{Component Evaluation}
\label{sec:component-evaluation}

We evaluate the impact of several design decisions on \querysmt's performance by rerunning \querysmt's evaluation subject to various modifications. Each row in \Cref{tbl:configuration-results} corresponds to a re-evaluation of \querysmt with one component of \querysmt's default implementation disabled or altered. ``Premises'' refer to facts recommended by premise selection (these are tailored to the current problem), while ``theory lemmas'' refer to the small set of properties about integers and natural numbers that \cvc5's hints aren't expected to capture (these are fixed across all problems). For more information on the set of theory lemmas used by \querysmt, either in the default configuration or in one of the modified configurations, see \Cref{sec:casting-facts}.

\begin{table}[ht]
\begin{center}
\caption{Benchmark theorems solved by \querysmt subject to modifications}
\label{tbl:configuration-results}
\begin{tabular}{|l|c|c|c|}
\hline
& Int Theorems & Nat Theorems & List Theorems\\
\hline
Default & 840 & 892 & 749 \\
\hline
No Hints & 472 & 627 & 708 \\
\hline
No Theory Lemmas & 697 & 625 & 736 \\
\hline
No Set of Support Strategy & 721 & 613 & 434 \\
\hline
Hints not in Set of Support & 762 & 659 & 733 \\
\hline
Premises not in Set of Support & 807 & 838 & 480 \\
\hline
Only Unit Theory Lemmas & 821 & 847 & 735 \\
\hline
With AC Theory Lemmas & 828 & 883 & 736 \\
\hline
\end{tabular}
\end{center}
\end{table}

In \Cref{sec:results}, we discuss the impact of removing hints from \querysmt (the ``No Hints'' row of \Cref{tbl:configuration-results} corresponds with the \querysmtminus row of \Cref{tbl:benchmark-results}). Here, we comment on other design decisions less central to the overall narrative.

The inclusion of theory lemmas appears to play an important role in allowing \duper to effectively use \cvc5's hints. From the fact that more Int theorems are solved with hints but no theory lemmas than with theory lemmas but no hints, we infer that hints retain some utility even in the absence of theory lemmas. But from the fact that \querysmt performs nearly identically on Nat theorems in the ``No Hints'' and ``No Theory Lemmas'' configurations, we infer that the theory lemmas are essential in practice for applying \cvc5's integer theory hints to problems about natural numbers.

The set of support strategy added to \duper appears to play an important role in limiting the theory lemmas' explosive behavior. We see this most clearly in the performance of the ``No Set of Support Strategy'' configuration on List theorems. In \querysmt's default setting, the only facts excluded from the set of support are theory lemmas, so the only facts impacted by disabling \duper's set of support strategy are these theory lemmas. From the significant drop in performance from \querysmt's default setting to ``No Set of Support Strategy'' configuration on List theorems, we infer that the theory lemmas exhibit explosive behavior that can effectively be reigned in by excluding them from the set of support. Conversely, from the fact that performance improves when hints and premises are included in the set of support, we infer that hints and premises do not generally exhibit this explosive behavior (or if they do, their relevance to the target goal outweighs the performance hit caused by any explosive behavior).

The exact set of theory lemmas included do not appear to have a large impact on \querysmt's performance. This is somewhat surprising given that \cvc5's normalization of AC operators is not captured by the hints it produces (see \Cref{sec:hint-generation}). \Cref{tbl:configuration-results} indicates that \querysmt does better in the default setting than in a setting which includes AC theory lemmas, but based on our anecdotal experience working with the tool, we suspect that adding or removing AC lemmas is simply introducing noise that by coincidence causes \duper to do better or worse on some problems. Equipping \querysmt with better instrumentation for tracking how \cvc5 normalizes AC operators may yield better results than trying to handle AC transformations with a static set of theory lemmas.
\section{Theory Lemmas Used by \querysmt}
\label{sec:casting-facts}

\Cref{tbl:casting-facts} contains a list of theory lemmas that are always provided to \duper (but excluded from \duper's set of support). These lemmas are intended to capture properties about integers and natural numbers that \cvc5's hints are not expected to cover. Most either relate to casting between Lean's \texttt{Nat} and \texttt{Int} types or equating terms that \cvc5 views as equivalent. The ``D'', ``Unit'', and ``AC'' columns indicate whether the lemma is included in \querysmt's default implementation, ``Only Unit Theory Lemmas'' implementation, and ``With AC Theory Lemmas'' implementation respectively. Unit theory lemmas are theory lemmas that resolve to unit clauses when processed by \duper, and AC theory lemmas are lemmas that assert associativity or commutativity properties.

\begin{table}[ht]
\begin{center}
\caption{Theory Lemmas used by \querysmt}
\label{tbl:casting-facts}
\parbox{.6\linewidth}{
\centering
\begin{tabular}{|l|c|c|c|}
\hline
Theory Lemma & D & Unit & AC\\
\hline
\texttt{Nat.zero\_le} & Y & Y & Y\\
\hline
\texttt{Int.natCast\_nonneg} & Y & Y & Y\\
\hline
\texttt{ge\_iff\_le} & Y & Y & Y\\
\hline
\texttt{gt\_iff\_lt} & Y & Y & Y\\
\hline
\texttt{lt\_iff\_not\_ge} & Y & Y & Y\\
\hline
\texttt{le\_iff\_lt\_or\_eq} & Y & N & Y\\
\hline
\texttt{Int.ofNat\_inj} & Y & Y & Y\\
\hline
\texttt{Int.natAbs\_natCast} & Y & Y & Y\\
\hline
\texttt{Int.natAbs\_eq} & Y & N & Y\\
\hline
\texttt{Int.natAbs\_eq\_natAbs\_iff} & Y & N & Y\\
\hline
\texttt{Int.ofNat\_le} & Y & Y & Y\\
\hline
\texttt{Int.ofNat\_lt} & Y & Y & Y\\
\hline
\texttt{Int.ofNat\_eq\_coe} & Y & Y & Y\\
\hline
\texttt{Int.zero\_sub} & Y & Y & Y\\
\hline
\texttt{Int.natAbs\_of\_nonneg} & Y & N & Y\\
\hline
\texttt{Int.ofNat\_natAbs\_of\_nonpos} & Y & N & Y\\
\hline
\texttt{Int.nonpos\_of\_neg\_nonneg} & Y & N & Y\\
\hline
\texttt{Int.nonneg\_of\_neg\_nonpos} & Y & N & Y\\
\hline
\end{tabular}
}
\parbox{.37\linewidth}{
\centering
\begin{tabular}{|l|c|c|c|}
\hline
Theory Lemma & D & Unit & AC\\
\hline
\texttt{Int.natCast\_add} & Y & Y & Y\\
\hline
\texttt{Int.natCast\_mul} & Y & Y & Y\\
\hline
\texttt{Int.natAbs\_mul} & Y & Y & Y\\
\hline
\texttt{Int.natCast\_one} & Y & Y & Y\\
\hline
\texttt{Int.natCast\_zero} & Y & Y & Y\\
\hline
\texttt{Int.natAbs\_zero} & Y & Y & Y\\
\hline
\texttt{Int.natAbs\_one} & Y & Y & Y\\
\hline
\texttt{Int.ofNat\_zero} & Y & Y & Y\\
\hline
\texttt{Int.ofNat\_one} & Y & Y & Y\\
\hline
\texttt{Int.mul\_assoc} & N & N & Y\\
\hline
\texttt{Int.mul\_comm} & N & N & Y\\
\hline
\texttt{Int.add\_assoc} & N & N & Y\\
\hline
\texttt{Int.add\_comm} & N & N & Y\\
\hline
\texttt{Nat.mul\_assoc} & N & N & Y\\
\hline
\texttt{Nat.mul\_comm} & N & N & Y\\
\hline
\texttt{Nat.add\_assoc} & N & N & Y\\
\hline
\texttt{Nat.add\_comm} & N & N & Y\\
\hline
\end{tabular}
}
\end{center}
\end{table}


\end{document}